\documentclass[a4paper,11pt]{article}
\usepackage[latin1]{inputenc}
\usepackage{amsmath}
\usepackage{amssymb}

\usepackage[affil-it]{authblk}

\title{On the Topological Complexity of $\om$-Languages of   Non-Deterministic Petri Nets }

\author{Olivier Finkel}
\affil{{\it Equipe de Logique Math\'ematique} \\ Institut de Math\'ematiques de Jussieu --- Paris Rive Gauche 
  \\ CNRS et  Universit\'e Paris Diderot Paris 7 \\
UFR de Mathématiques case 7012
\\75205 Paris Cedex 13,  France.\\ finkel@math.univ-paris-diderot.fr 
}

\author{Micha{\l} Skrzypczak\thanks{This author has been supported by National Science Centre grant no. DEC-2012/05/N/ST6/03254.}}
\affil{Institute of Informatics\\ University of Warsaw\\ Banacha 2\\ 02-097 Warsaw, Poland.\\ mskrzypczak@mimuw.edu.pl}

\usepackage {tikz}

\usetikzlibrary{positioning}
\usetikzlibrary{decorations.pathreplacing}
\usetikzlibrary{automata,positioning}
\usetikzlibrary{decorations.pathmorphing}
\usetikzlibrary{decorations.markings}
\usetikzlibrary{arrows}

\date{}
\begin{document}

\newcommand{\inc}{+1}
\newcommand{\eps}{0}
\newcommand{\dec}{-1}

\newcommand{\comment}[1]{}

\newtheorem{The}{Theorem}[section]
\newtheorem{Pro}[The]{Proposition}
\newtheorem{Deff}[The]{Definition}
\newtheorem{Lem}[The]{Lemma}
\newtheorem{Fac}[The]{Fact}
\newtheorem{Rem}[The]{Remark}
\newtheorem{Exa}[The]{Example}
\newtheorem{Cor}[The]{Corollary}
\newtheorem{Not}[The]{Notation}

\newcommand{\fa}{\forall}
\newcommand{\Ga}{\mathsf\Gamma}
\newcommand{\Gas}{\Ga^\star}
\newcommand{\Gao}{\Ga^\omega}

\newcommand{\Si}{\mathsf\Sigma}
\newcommand{\Sis}{\Si^\star}
\newcommand{\Sio}{\Si^\omega}
\newcommand{\ra}{\rightarrow}
\newcommand{\hs}{\hspace{12mm}

\noi}
\newcommand{\lra}{\leftrightarrow}
\newcommand{\la}{language}
\newcommand{\ite}{\item}
\newcommand{\Lp}{L(\varphi)}
\newcommand{\abs}{\{a, b\}^\star}
\newcommand{\abcs}{\{a, b, c \}^\star}
\newcommand{\ol}{ $\omega$-language}
\newcommand{\orl}{ $\omega$-regular language}
\newcommand{\om}{\omega}
\newcommand{\nl}{\newline}
\newcommand{\noi}{\noindent}
\newcommand{\tla}{\twoheadleftarrow}
\newcommand{\de}{deterministic }
\newcommand{\proo}{\noi {\bf Proof.} }
\newcommand {\ep}{\hfill $\square$}
\renewcommand{\thefootnote}{\star{footnote}} 
\newcommand{\trans}[1]{\stackrel{#1}{\rightarrow}}
\newcommand{\dom}{\operatorname{Dom}}
\newcommand{\im}{\operatorname{Im}}

\newcommand{\N}{\mathbb{N}}

\maketitle 

\begin{abstract}
\noi We  show that there are  ${\bf \Sigma}_3^0$-complete languages of infinite words accepted by non-deterministic Petri nets with B\"uchi acceptance 
condition, or equivalently by 
B\"uchi blind counter automata. This shows that $\om$-languages accepted by  non-deterministic Petri nets are topologically more complex than those 
accepted by deterministic Petri nets. 
\end{abstract}

\noi {\bf Keywords.} Languages of infinite words; Petri nets; B\"uchi acceptance condition; Cantor space; 
topological complexity; Borel hierarchy, complete sets.

\section{Introduction}

The languages of infinite words, also called  $\omega$-languages,  accepted by finite automata were first studied by B\"uchi 
to prove the decidability of the monadic second order theory of one successor
over the integers.  Since then regular $\omega$-languages have been much studied and  used  for specification and verification 
of non-terminating systems, 
see \cite{Thomas90,Staiger97,PerrinPin} for many results and references. 
The acceptance of infinite words by other   finite machines, like 
pushdown automata, counter automata, Petri nets, Turing machines, \ldots, with various acceptance conditions, 
 has also been considered,  see  \cite{Staiger97,eh,CG78b,Fin-survey}. 

Since the set $\Si^\omega$ of infinite words over a finite alphabet $\Si$ is naturally equipped with the Cantor topology, a
 way to study the complexity of   languages of infinite words  
accepted by finite machines          is to study    their      topological complexity       and firstly 
to locate them with regard to 
the Borel and the projective hierarchies \cite{Thomas90,Simonnet92,eh,LescowThomas,Staiger97,Selivanov08m,Selivanov08t}.  

Every $\omega$-language accepted by a deterministic B\"uchi automaton is a ${\bf \Pi}^0_2$-set. 
On the other hand   it follows from Mac Naughton's Theorem that 
an $\omega$-language accepted by a non-deterministic B\"uchi  (or   Muller)  automaton   is also accepted by a   deterministic
Muller   automaton, and thus is a boolean combination of $\omega$-languages accepted by  deterministic B\"uchi automata. Therefore 
every $\omega$-language accepted by a non-deterministic B\"uchi  (or   Muller)  automaton is a 
${\bf \Delta}^0_3$-set. 
In a similar way,  every $\omega$-language accepted by a deterministic Muller Turing machine, and thus also by any 
Muller deterministic finite machine is a ${\bf \Delta}^0_3$-set, \cite{eh,Staiger97}. 

We consider here acceptance of infinite words by Petri nets or equivalently by  (partially) blind counter automata. 
Petri nets are used for the description of distributed systems  \cite{esparza1998decidability,rozenberg2004lectures,H-petrinet-diaz}, and  they may be defined as 
partially blind multicounter automata, as explained in \cite{valk1983infinite,eh,Gre78}. 
In order to get a partially blind multicounter automaton which accepts the 
same language as a given Petri net, one can distinguish between the 
places of a Petri net by dividing them  into the bounded ones (the number of tokens 
in such a place at any time is uniformly bounded) and the unbounded ones. Then 
each unbounded place may be seen as a partially blind counter, and the tokens in the bounded places 
determine the state of the partially blind multicounter automaton. The transitions of the Petri net may then be seen as the finite control of the 
partially blind multicounter automaton and the labels of these transitions are then the input symbols. 

The infinite behavior of Petri nets was first studied by 
Valk \cite{valk1983infinite} and  by Carstensen in the case of deterministic Petri nets  \cite{carstensen1988infinite}.  

On one side the topological complexity of     $\om$-languages of  deterministic Petri nets is completely determined. They  are  ${\bf \Delta}^0_3$-sets 
and their Wadge hierarchy, which is a great refinement of the Borel hierarchy, defined via reductions by continuous functions, 
 has been determined in \cite{Fin01csl,DFR2,DFR3}; its length is the ordinal $\om^{\om^2}$.  

On the other side, nothing was known about the  topological complexity of     $\om$-languages of  non-deterministic Petri nets.  
We show that there exist  ${\bf \Sigma}_3^0$-complete, hence non  ${\bf \Delta}_3^0$,   $\om$-languages accepted by one-blind-counter B\"uchi automata. 
Notice that it was proved in \cite{Fin-mscs06} that $\om$-languages accepted by (non-blind) one-counter 
B\"uchi  automata have the same topological complexity as 
$\om$-languages of Turing machines, but the non-blindness of the counter was essential 
in the proof since the ability to use the zero-test of the counter was important.

This provides the first result on the topological complexity of $\om$-languages of  non-deterministic Petri nets 
and shows that there exist some $\om$-languages accepted by  non-deterministic Petri nets, and even by  one-blind-counter B\"uchi automata, 
which are topologically more complex than those 
accepted by deterministic Petri nets.

\section{Basic notions}

\hs We assume   the reader to be familiar with the theory of formal ($\om$)-languages, see   
\cite{Thomas90,Staiger97}.
  
When $\Si$ is a countable alphabet, a {\it non-empty finite word} over $\Si$ is any 
sequence $x=a_1\ldots a_k$, where $a_i\in\Si$ 
for $i=1,\ldots ,k$ , and  $k$ is an integer $\geq 1$.  
 $\Sis$  is the {\it set of finite words} (including the empty word $\varepsilon$) over $\Si$.
 
The {\it first infinite ordinal} is $\om$.
 An $\om$-{\it word} over $\Si$ is an $\om$ -sequence $a_1 \ldots a_n \ldots$, where for all 
integers $ i\geq 1$, ~
$a_i \in\Si$.  When $\sigma$ is an $\om$-word over $\Si$, we write
 $\sigma =\sigma(1)\sigma(2)\ldots \sigma(n) \ldots $,  where for all $i$,~ $\sigma(i)\in \Si$, and 
$\sigma[n] =\sigma(1)\sigma(2)\ldots \sigma(n)$.

 The concatenation product of two finite words $u$ and $v$ is 
denoted $u\cdot v$ and sometimes just $uv$. This product is extended to the product of a 
finite word $u$ and an $\om$-word $v$: the infinite word $u\cdot v$ is then the $\om$-word such that:
~~  $(u\cdot v)(k)=u(k)$  if $k\leq |u|$ , and 
 $(u\cdot v)(k)=v(k-|u|)$  if $k>|u|$.
  
 The {\it set of } $\om$-{\it words} over  the alphabet $\Si$ is denoted by $\Si^\om$.
An  $\om$-{\it language} over an alphabet $\Si$ is a subset of  $\Si^\om$.  

\hs A  blind  multicounter automaton is a finite automaton equipped 
with a finite number ($k$) of  blind (sometimes called partially blind, as in \cite{Gre78}) counters. The content of any such counter 
is a non-negative  integer. A  counter is said to be  blind when the 
multicounter automaton cannot test whether the content of the counter is zero. 
This means that  if a transition of the machine is enabled  when the content of a counter 
is zero then the same transition is also enabled when the content of the same counter 
is a non-zero integer.

We now give the definition of   a B\"uchi $1$-blind-counter automaton. Notice that we consider here only 
real time automata, i.e., without  $\varepsilon$-transitions. 

\begin{Deff} 
A (real time) B\"uchi $1$-blind-counter automaton  is a 5-tuple 
$\mathcal{A}=(Q,\Si, \Delta, q_0, F)$,  where $Q$ 
is a finite set of states, $\Si$ is a finite input alphabet, 
 $q_0\in Q$ is the initial state, 
 the transition relation $\Delta$ is a subset of  
$Q \times \Si \times \{0, 1\}  \times  Q \times \{0, 1, -1\}$, and $F\subseteq Q$ is the set of accepting states.

If  the  automaton   $\mathcal{A}$  is in state $q$, and  
$c \in \mathbb{N}$ is the content of  
the counter $\mathcal{C}$, then 
the  configuration (or global state)
 of $\mathcal{A}$ is the  pair $(q, c)$.

Given any $a\in \Si$, any $q, q' \in Q$, and any $c \in \mathbb{N}$, if both 
 $\Delta(q, a, i, q', j)$, and $(c \geq 1  \Rightarrow i=1)$  and  $(c=0 \Rightarrow ( i=0  \mbox{ and } j\in\{0,1\} ) )$ holds, then 
 we write:  ~~  $a: (q, c)\mapsto_{\mathcal{A}} (q', c+j)$. 

\comment{
Thus we see that the transition relation must verify:
\nl if $\Delta(q, a, i, q', j)$, and  $i=0$ holds  
 then we must have $j=0$ or $j=1$. 
}

Moreover the  
counter of $\mathcal{A}$ is blind, {\em i.e.}, if $\Delta(q, a, i, q', j)$ holds, and  $i=0$ 
 then $\Delta(q, a, i', q', j)$ holds also for  $i'=1$.

\hs
Let $x =a_1a_2 \ldots a_n \ldots $ be an $\omega$-word over $\Si$. 
An $\omega$-sequence of configurations $\rho=(q_i, c_{i})_{i \geq 1}$ is called 
a  run of $\mathcal{A}$ on $x$ if and only if
\begin{itemize}
\item $(q_1, c_{1})=(q_0, 0)$, and 
\vspace{1ex}
\item   
 $a_i: (q_i, c_{i})\mapsto_{\mathcal{A}}
(q_{i+1},  c_{i+1})$ (for all $1\leq i$).
\end{itemize}

We denote  $In(\rho)$  the set of all the states visited infinitely
 often during the  run $\rho$.
The automaton    $\mathcal{A}$    accepts $x$ 
 if there is an infinite run $\rho$ of  $\mathcal{A}$ on $x$   such that    $In(\rho)\cap F\neq\emptyset$. 

The $\om$-language accepted by  $\mathcal{A}$  is  the set  $L(\mathcal{A})$ of $\om$-words accepted by $\mathcal{A}$. 

\end{Deff}

We assume the reader to be familiar with basic notions of topology which
may be found in \cite{Moschovakis80,LescowThomas,Kechris94,Staiger97,PerrinPin}.
If $X$ is a countable alphabet containing at least two letters, then the set  $X^\om$ of infinite words over $X$ may be equipped with the product topology of the discrete topology on $X$. 
This topology is induced by  a natural metric  which is called the {\it prefix metric} and defined as follows. For $u, v \in X^\om$ and 
$u\neq v$ let $\delta(u, v)=2^{-l_{\mathrm{pref}(u,v)}}$ where $l_{\mathrm{pref}(u,v)}$ 
 is the first integer $n$
such that the $(n+1)^{st}$ letter of $u$ is different from the $(n+1)^{st}$ letter of $v$. 

If $X$ is finite  then $X^\om$ is a Cantor space and if $X$ is countably infinite then $X^\om$ is homeomorphic to the Baire space $\om^\om$. 
 The open sets of $X^\om$ are the sets in the form $W\cdot X^\om$, where $W\subseteq X^\star$.

The classes ${\bf \Sigma}_n^0$ and ${\bf \Pi}_n^0$ of the Borel Hierarchy
 on the topological space $X^\om$  are defined as follows:
${\bf \Sigma}^0_1$ is the class of open sets of $X^\om$, ${\bf \Pi}^0_1$ is the class of closed sets (i.e. complements of open ones) 
of $X^\om$.
 And for any integer $n\geq 1$:
${\bf \Sigma}^0_{n+1}$   is the class of countable unions 
of ${\bf \Pi}^0_n$-subsets of  $X^\om$, and 
 ${\bf \Pi}^0_{n+1}$ is the class of countable intersections of 
${\bf \Sigma}^0_n$-subsets of $X^\om$.
The Borel Hierarchy is also defined for transfinite levels, but we shall not 
need them in the present study. 

Recall now  the notion of completeness with regard to reduction by continuous functions. 
For an integer $n\geq 1$, a set $F\subseteq X^\om$ is said to be 
a ${\bf \Sigma}^0_n$  (respectively,  ${\bf \Pi}^0_n$)-hard set 
iff for any set $E\subseteq Y^\om$  (with $Y$ a countable alphabet): $E\in {\bf \Sigma}^0_n$ (respectively,  $E\in {\bf \Pi}^0_n$) 
implies that there exists a continuous function $f: Y^\om \ra X^\om$ such that $E = f^{-1}(F)$.  If the set $F$ is 
${\bf \Sigma}^0_n$  (respectively,  ${\bf \Pi}^0_n$)-hard {\it and } belongs to the class ${\bf \Sigma}^0_n$  
(respectively,  ${\bf \Pi}^0_n$) then it is said to be 
${\bf \Sigma}^0_n$  (respectively,  ${\bf \Pi}^0_n$)-complete. 


\section{Topological complexity of  Petri nets $\om$-languages}

We now state and prove our main result. 

\begin{The}
There exists a ${\bf \Sigma}^0_3$-complete $\om$-language accepted by a  B\"uchi one-blind-counter automaton. 
\end{The}

The rest of this section is devoted to showing this result. First, we describe the construction of an automaton $\mathcal{A}$ recognising a ${\bf \Sigma}^0_3$-hard language. 
Then, in Lemma~\ref{lem:upper}, we show that the language recognised by $\mathcal{A}$ belongs to ${\bf \Sigma}^0_3$.

Let us recall an example of a ${\bf \Pi}^0_3$-complete subset $C_3$ of the Baire space given in \cite[page 180]{Kechris94}. 

$$C_3= \{ x\in \om^\om \mid \lim_{n} x(n) = \infty \}$$

\noi It follows that the set 
$$D_3= \{ x\in \om^\om \mid \liminf x(n) < \infty \}$$

\noi  is ${\bf \Sigma}^0_3$-complete. 

\hs Notice that we have 

$$D_3= \{ x\in \om^\om \mid \exists N  ~\exists^\infty i ~ (x(i) \leq N) \}=\{ x\in \om^\om \mid \exists N  ~  \fa p ~ \exists  i >p ~ (x(i) \leq N) \}$$

\noi We now define the following coding of infinite sequences of integers by infinite words over the alphabet $\Si=\{a, b\}$. 
For  $x=(m_i)_{i\geq 0} \in \om^\om$ we set $n_i=m_i + 1$ for each $i\geq 0$, and 
$$\Phi (x) = a^{n_0}\cdot b^{n_0}\cdot a^{n_1}\cdot b^{n_1}\cdots a^{n_i}\cdot b^{n_i}\cdots $$
\noi It is clear that this defines a {\it continuous } injective mapping  $\Phi: \om^\om \ra \{a, b\}^\om$. 
We are going to  show that there exists a one-blind-counter automaton $\mathcal{A}$,  reading $\om$-words over $\Si$, such that 
$L(\mathcal{A})\cap  \Phi (\om^\om) =  \Phi (D_3)$, what is equivalent to
\begin{equation}
\fa x \in \om^\om ~~ x\in D_3  \Longleftrightarrow  \Phi (x) \in L(\mathcal{A})\label{eq:reduction}
\end{equation}
\noi It implies that $L(\mathcal{A})$ is ${\bf \Sigma}^0_3$-hard, since $D_3$ is ${\bf \Sigma}^0_3$-complete.

\comment{
\begin{figure}
\begin{tikzpicture}[shorten >=1pt,node distance=1.5cm,on grid,auto]

\node[state, initial] (qI) at (-1,2) {$q_I$};
\node[state] (qA) at (0,0) {$q_a$};
\node[state, accepting] (qF) at (3,1) {$q_F$};
\node[state] (qB) at (6,0) {$q_b$};
\node[state] (qiA) at (4.5,-4) {$p_a$};
\node[state] (qiB) at (0,-4) {$p_b$};

\path[->] 
    (qI) edge [loop above] node {$a,b: \inc$} (qI)
    (qI) edge [swap] node {$a: \dec$} (qA)
    (qA) edge [loop left] node {$a: \dec$} (qA)
    (qA) edge [bend left] node {$b: \inc$} (qF)
    (qF) edge [bend left=0] node {$b: \inc$} (qB)
    (qB) edge [loop right] node {$b: \inc$} (qB)
    (qF) edge [bend left=10] node {$a$} (qiA)
    (qB) edge node {$a$} (qiA)
    (qF) edge node {$a:\dec$} (qA)
    (qB) edge [bend left=10] node {$a:\dec$} (qA)
    (qiA) edge [loop below] node {$a$} (qiA)
    (qiA) edge [bend left] node {$b$} (qiB)
    (qiB) edge [loop below] node {$b$} (qiB)
    (qiB) edge [bend left] node {$a$} (qiA)
    (qiB) edge node {$a:\dec$} (qA);
\end{tikzpicture}
\caption{The automaton $\mathcal{A}$}
\label{fig:A}
\end{figure}
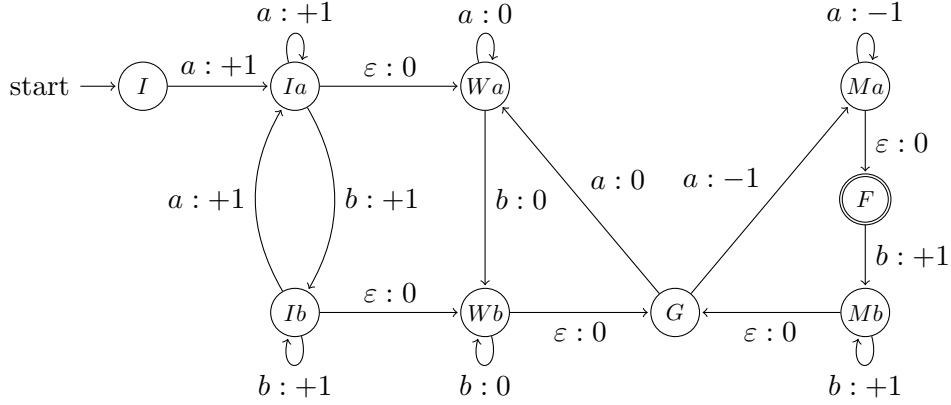}

We start with a formal definition of $\mathcal{A}$ as depicted on Figure~\ref{fig:A}: the initial state is denoted $I$ and the unique accepting state is denoted $F$. An edge of the form $q\overset{a:j}{\longrightarrow} q'$ denotes the pair of transitions $(q,a,0,q',j)$ and $(q,a,1,q',j)$.

For the sake of readability we use $\varepsilon$-transitions in $\mathcal{A}$. They can be eliminated in the standard way as there is no loop with only $\varepsilon$-transitions and the $\varepsilon$-transitions do not modify the counter.

\begin{figure}[h]
\begin{tikzpicture}[shorten >=1pt,node distance=1.5cm,on grid,auto]

\tikzstyle{st}=[state,minimum size=0.8cm,inner sep=0pt,scale=0.8]

\node[st, initial] (qI) at (-1,0) {${}I$};
\node[st] (Ia) at (1,0) {${}{Ia}$};
\node[st] (Ib) at (1,-3) {${}{Ib}$};

\node[st] (Wa) at (3.5,0) {${}{Wa}$};
\node[st] (Wb) at (3.5,-3) {${}{Wb}$};

\node[st] (G) at (6,-3) {${}{G}$};

\node[st] (Ma) at (8.5,0) {${}{Ma}$};
\node[st, accepting] (Mf) at (8.5,-1.5) {${}{F}$};
\node[st] (Mb) at (8.5,-3) {${}{Mb}$};

\path[->] 
    (qI) edge [] node {$a: \inc$} (Ia)
    (Ia) edge [loop above] node {$a: \inc$} (Ia)
    (Ia) edge [bend left] node {$b: \inc$} (Ib)
    (Ib) edge [loop below] node {$b: \inc$} (Ib)
    (Ib) edge [bend left] node {$a: \inc$} (Ia)
    
    (Ia) edge [] node {$\varepsilon: \eps$} (Wa)
    (Ib) edge [] node {$\varepsilon: \eps$} (Wb)
    (Wa) edge [loop above] node {$a: \eps$} (Wa)
    (Wa) edge [] node {$b: \eps$} (Wb)
    (Wb) edge [loop below] node {$b: \eps$} (Wb)
    (Wb) edge [swap] node {$\varepsilon: \eps$} (G)

    (G) edge [] node {$a: \dec$} (Ma)            
    (Ma) edge [loop above] node {$a: \dec$} (Ma)
    (Ma) edge [] node {$\varepsilon: \eps$} (Mf)
    (Mf) edge [] node {$b: \inc$} (Mb)  
    (Mb) edge [loop below] node {$b: \inc$} (Mb)
    (Mb) edge [] node {$\varepsilon: \eps$} (G)
    
    (G) edge [swap] node {$a: \eps$} (Wa);
\end{tikzpicture}
\caption{The automaton $\mathcal{A}$}
\label{fig:A}
\end{figure}

The automaton is designed in such a way to remember the last letter read: in the states $Ia$, $Wa$, $Ma$, and $F$ it is $a$ and in the states $Ib$, $Wb$, $G$, and $Mb$ it is $b$.

Let $Z$ be the set of words $w\in\Si^\omega$ that contain infinitely many letters $a$ and $b$ and start from $a$. Clearly $L(\mathcal{A})\subseteq Z$. Observe that if $w\in Z$ is an infinite word then it can be uniquely decomposed as $w=a^{n_0}b^{k_0}a^{n_1}b^{k_1}\ldots$ with $n_i,k_i>0$. A subword $a^{n_i} b^{k_i}$ of $w$ in the above decomposition is called a \emph{block of $w$}.

Any accepting run of $\mathcal{A}$ on a word $w$ can be divided into three stages:
\begin{enumerate}
\item in the states $I$, $Ia$, and $Ib$ automaton $\mathcal{A}$ reads first $N$ letters of $w$ and increments the counter,
\item in the states $Wa$ and $Wb$ automaton $\mathcal{A}$ reads the rest of the current block,
\item infinitely often automaton is in $G$ at the beginning of some block $a^n b^k$ and decides either to read it without changing the counter (states $Wa,Wb$) or to:
\begin{itemize}
\item decrease the counter on $a^n$ in the state $Ma$, 
\item visit once the accepting state $F$,
\item increase the counter on $b^k$ in $Mb$.
\end{itemize}
\end{enumerate}
Note that $\mathcal{A}$ can read a block $a^nb^k$ using $Ma$, $F$, and $Mb$ only if the counter value at the beginning of this block is at least $n$.

By the definition of the automaton the language recognised by $\mathcal{A}$ is the set of all words of the form $w=a^{n_0}b^{k_0}a^{n_1}b^{k_1}\ldots$ with $n_i,k_i>0$ such that for some $N\in\N$ and a set $I\subseteq \N$ we have:
\begin{itemize}
\item $I$ is infinite,
\item the block $a^{n_i} b^{k_i}$ for $i=\min(I)$ starts after the $N$'th letter of $w$,
\item for every $i\in I$ we have
\begin{equation}
n_i\leq N + \sum_{j<i\wedge j\in I} k_j-n_j.\label{eq:defL}
\end{equation}
\end{itemize}

Now we want to prove (\ref{eq:reduction}). Let $x=(m_i)_{i\geq 0}$ be an element of the Baire space. Observe that $w=\Phi(x)$ has the form $a^{m_0+1}b^{m_0+1}a^{m_1+1}b^{m_1+1}\ldots$. Therefore, in the above conditions $k_j=n_j$ and (\ref{eq:defL}) takes the form
\[n_i \leq N\quad\text{with}\quad n_i=m_i+1.\]

Now, to prove (\ref{eq:reduction}) it is enough to observe that the following conditions are equivalent
\begin{itemize}
\item $\mathcal{A}$ accepts $\Phi(x)$
\item there exists $N$ such that there are infinitely many $i$ with $m_i\leq N$
\item $x\in D_3$.
\end{itemize}

\comment{
Now we explain informally the behaviour of the automaton $\mathcal{A}$ on a code $\Phi (x)$ of an element $x=(m_i)_{i\geq 0}$ of the Baire space. 

At the beginning of a run of $\mathcal{A}$ on $\Phi (x)$ the counter of the automaton is increased non-deterministically on states $I$, $Ia$, and $Ib$ until some value $k\geq 1$ is reached. Then the automaton waits until the end of the current block in states $Wa$ and $Wb$.

Now the automaton has to check that there are infinitely many integers $i$ such that $n_i$ is smaller than $k$. 
Again this is done in a non-deterministic way. When reading the blocks $a^{n_i}\cdot b^{n_i}$ of  $\Phi (x)$, the automaton chooses infinitely often to check 
that $n_i \leq k$. To do this it guesses at some time that a given block $a^{n_i}\cdot b^{n_i}$ is a good one by moving from $G$ to $Ma$. Then it
first decreases its counter of $1$ for each letter $a$ read; then if the automaton is not blocked this implies that 
$n_i \leq k$, and otherwise the automaton is blocked and the run cannot be an accepting one. Next when reading the letters $b$ forming the block $b^{n_i}$ 
the automaton increases by $1$ its counter for each letter $b$ read; then after having read the block $a^{n_i}\cdot b^{n_i}$ of  $\Phi (x)$ 
the value of the counter is again $k$ and the automaton can repeat this operation for another block $a^{n_j}\cdot b^{n_j}$ of  $\Phi (x)$. 
The B\"uchi acceptance condition is used to check that this operation is done {\it infinitely often}. Therefore, we obtain (\ref{eq:reduction}).}


\begin{Lem}\label{lem:upper}
The language $L(\mathcal{A})$ is a ${\bf \Sigma}^0_3$ set.
\end{Lem}

\proo
First note that the set $Z$ defined above is a ${\bf \Pi}^0_2$ set, so we can restrict our attention only to words in $Z$.

Let $\rho$ be a run of $\mathcal{A}$ on a given infinite word $w=a^{n_0}b^{k_0}a^{n_1}b^{k_1}\ldots$ with $n_i,k_i>0$. We say that $\rho$ \emph{uses} a block $a^{n_i}b^{k_i}$ in $w$ if $\mathcal{A}$ decreases its counter on $a^{n_i}$ in the state $Ma$ and then increases it on $b^{k_i}$ in the state $Mb$. We say that a block $a^{n_i}b^{k_i}$ is \emph{positive} if $k_i\geq n_i$.

Now, fix a number $N$ (the guessed initial value of the counter). We describe how to inductively construct a run $\rho(N)$ of $\mathcal{A}$ on $w$. The run starts by increasing the counter $N$ times and waiting until the end of the current block. Then, when reaching the first letter of some block $a^nb^k$ with counter value $c$ in the state $G$, the following cases are possible:
\begin{itemize}
\item the block $a^nb^k$ is not positive then $\rho(N)$ does not use it (moves to $Wa$),
\item the block is positive but $n>c$ then $\rho(N)$ does not use it (moves to $Wa$),
\item the block is positive and $n\leq c$ then $\rho(N)$ uses it (moves to $Ma$).
\end{itemize}

The following fact describes the crucial property of the run $\rho(N)$.

\begin{Fac}\label{ft:implication}
If $\rho$ is an accepting run of $\mathcal{A}$ on $w$ that starts with $N$ increments then $\rho(N)$ is also accepting.
\end{Fac}

\proo
Observe that if $\rho$ is accepting then it uses infinitely many positive blocks --- otherwise only finitely many non-positive ones can be used.

Now, inductively show that $\rho(N)$ has value of the counter at least equal to the value of $\rho$. 
In particular, whenever $\rho$ uses some positive block then $\rho(N)$ also uses it. Therefore, $\rho(N)$ uses infinitely many blocks and accepts.
\ep 

Observe that given a number $N$ the condition ``$\rho(N)$ uses the $i$'th block $a^{n_i}b^{k_i}$" is an open property depending only on the initial segment of $w$ until the end of the block $a^{n_i}b^{k_i}$. Consider the following formula:
\[\varphi:=\exists_N\forall_i\exists_{j\geq i}\ \mbox{$\rho(N)$ uses the $j$'th block $a^{n_j}b^{k_j}$}.\]

This formula is a ${\bf \Sigma}^0_3$ formula. We claim that $\varphi$ defines the language $L(\mathcal A)$. Clearly, if a word $w$ satisfies $\varphi$ then, for the appropriate value of $N$, the run $\rho(N)$ is accepting. For the other direction, assume that $\rho$ is an accepting run of $\mathcal{A}$ on $w$. By Fact~\ref{ft:implication} we know that there is $N$ such that $\rho(N)$ is accepting. Therefore, $\varphi$ is satisfied on $w$.
\ep

\section{Concluding remarks}

We have proved that there are some ${\bf \Sigma}_3^0$-complete languages of infinite words accepted by  blind-counter B\"uchi automata. 
This provides the first results on the   topological complexity of     $\om$-languages of  non-deterministic Petri nets 
and shows that $\om$-languages accepted by  non-deterministic Petri nets are topologically more complex than those 
accepted by deterministic Petri nets. 
A natural question is now to completely determine the Borel and Wadge hierarchies of  $\om$-languages of  {\it non-deterministic}  Petri nets. 
The first question would be: ``are there non  ${\bf \Sigma}_3^0$-sets accepted by {\it non-deterministic}  Petri nets?".

\end{document}